\documentclass[%
reprint,
 amsmath,amssymb,
 aps,
]{revtex4-2}

\usepackage{graphicx}
\usepackage{dcolumn}
\usepackage{bm}

\newcommand{\ttorque}{$\overline{\tau}_{\varphi}$}
\newcommand{\torque}{$\tau_{\varphi}$}
\newcommand{\Ms}{$M_{\mathrm{S}}$}
\newcommand{\Msne}{$M_{\mathrm{S,NE}}$}
\newcommand{\Hs}{$H_{\mathrm{S}}$}
\newcommand{\Aex}{$A_{\mathrm{ex}}$}
\newcommand{\Bext}{$B_{\mathrm{ext}}$}
\newcommand{\sigmams}{$\sigma_{M_{\mathrm{S}}}$}
\begin{document}

\preprint{APS/123-QED}

\title{Control of vortex chirality in a symmetric ferromagnetic ring using ferromagnetic nanoelement}

\author{Uladzislau Makartsou}
 \email{ulamak@st.amu.edu.pl}
 \affiliation{Institute of Spintronics and Quantum Information, Faculty of Physics, Adam Mickiewicz University, Uniwersytetu Poznańskiego 2, 61-614 Poznań, Poland}
\author{Mathieu Moalic}
\affiliation{Institute of Spintronics and Quantum Information, Faculty of Physics, Adam Mickiewicz University, Uniwersytetu Poznańskiego 2, 61-614 Poznań, Poland}
\author{Mateusz Zelent}
\affiliation{Institute of Spintronics and Quantum Information, Faculty of Physics, Adam Mickiewicz University, Uniwersytetu Poznańskiego 2, 61-614 Poznań, Poland}
\author{Michal Mruczkiewicz}
\affiliation{Institute of Electrical Engineering, Slovak Academy of Sciences, Dubravska cesta 9, SK-841-04 Bratislava, Slovakia and Centre For Advanced Materials Application CEMEA, Slovak Academy of Sciences, Dubravska cesta 9, 845 11 Bratislava, Slovakia}
\author{Maciej Krawczyk}
 \email{krawczyk@amu.edu.pl}
\affiliation{Institute of Spintronics and Quantum Information, Faculty of Physics, Adam Mickiewicz University, Uniwersytetu Poznańskiego 2, 61-614 Poznań, Poland}

\date{\today}

\begin{abstract}
Controlling the vortex chirality in ferromagnetic nanodots and nanorings has been a topic
of investigation for the last few years. Many control methods have been proposed and it has been found that the control is related to the breaking of the circular symmetry. In this paper, we present a theoretical study demonstrating the control of chirality in ferromagnetic nanoring without directly breaking its symmetry, but instead by placing elongated ferromagnetic nanoelement inside the ring, Here, the stray magnetostatic field exerted by the asymmetrically  placed nanoelement determines the  movement of the domain walls upon remagnetization of the nanoring and the resulting chirality in the remanence. This approach allows the chirality of the vortex state to be controlled and also promises its control in a dense array of nanorings, thus suitable for spintronic and magnonic applications.

\end{abstract}
\maketitle

\section{\label{sec:intro}Introduction}
An advantage of the soft ferromagnetic disks and rings over the monodomain nanoparticles for the development of magnetic memory and reprogrammable logic devices is vortex chirality.\cite{Zhu2000,Klaui2003,Vaz2007,Ross2008,Zheng2017} The key factor determining the chirality of ferromagnetic nanorings (NRs) is the control of domain wall (DW) motion  \cite{Hayward2005, Mawass2017}. Understanding and control of the transition process from an onion state (OS) to a vortex state (VS) opens opportunities not only in the study and development of applications based on the static magnetization configuration and dynamic properties related to the chirality of NRs \cite{Podbielski2006,Neudecker2006,Montencello2008} but may also contribute to devices based on DW dynamics in arrays of NRs \cite{Rapoport2012,Dawidek2021}. The transition from the OS to the VS can be induced by reducing the in-plane magnetic field. This results in moving head-to-head (HTH) or tail-to-tail (TTT) DWs  perpendicular to the field direction towards either the left or right side of the ring. The direction of DWs movement is spontaneous and random, leading to the NR obtaining a flux-closure state with either clockwise (CW) or counterclockwise (CCW) chirality. This indicates that the DW motion and vortex chirality control presented in our paper can be applied to a wide variety of spintronic and magnonic applications. 

Chirality control usually involves breaking the circular symmetry of the NR, as a result of which the direction of movement of the DWs when interacting with an external magnetic field is determined by the energy difference between the asymmetric sides of the ring \cite{Zhu2006,Jung2006}. For instance, as the asymmetry is introduced as a de-centered ring, i.e., one side of the ring has a larger width than the other, the direction of the external field applied in the plane parallel to these sides, determines the VS chirality \cite{Saitoh2004,Huang2012,Schonke2020}. Similarly to VS in full disks,  deformation of edges, like notches, also allows to control chirality by causing energy splitting under in-plane applied magnetic fields \cite{Klaui2001,Tobik2015,Zheng2017}.

Our approach proposes abandoning the direct change in a ring symmetry in favour of its altering by the magnetization of an asymmetrically located single-domain ferromagnetic nanoelement (NE) with high shape anisotropy, placed inside the NR. Its interaction with HTH-TTT DWs can provide the desired control of vortex chirality.
The NE must have the following properties in order to prevent its magnetization from changing by external magnetic field and stray fields produced by NR: sufficiently strong anisotropy and sufficiently high magnetic moment. In numerical simulations, we found that an NE having a shape of ring segment  and located asymmetrically within the NR would be most suitable. 

 Chirality control by magnetostatic coupling with nanomagnets in ferromagnetic nanodisks has recently been implemented by placing rhombus elements near the disk edges \cite{Haldar2015}. Our approach  adds additional flexibility to the process of controlling chirality and motion of DWs. Placing NE inside the ring can allow to create a dense array of NRs \cite{Tadmor2012}, and even an array of overlapping rings \cite{Dawidek2021}, with controlled chirality. This is important in the context of interconnected rings, which have recently been demonstrated to perform well in terms of reservoir computing system \cite{Dawidek2021,Vidamour2022}.

\section{\label{sec:methods}Geometry and simulation method}
\begin{figure}[h]
\includegraphics[width=0.5\textwidth]{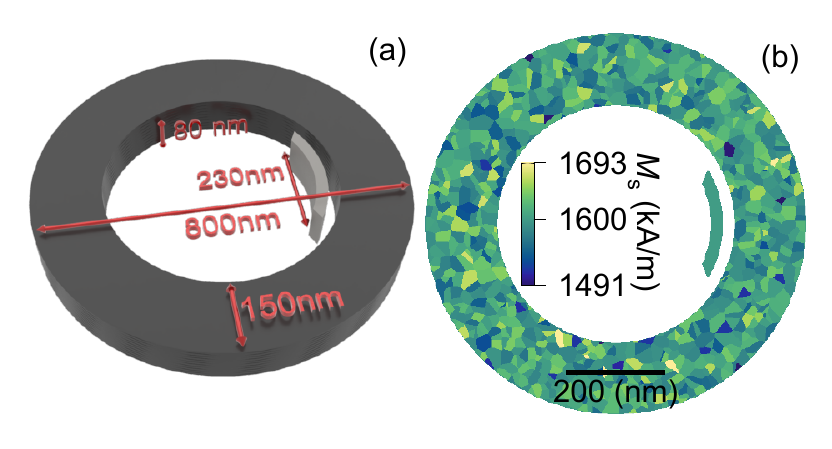}
\caption{\label{fig:structure}(a) Schematic representation of the system under consideration: the ferromagnetic nanoelement (NE) inside the ferromagnetic ring (NR). (b) The grains structure representation used in simulations.}
\end{figure}

We study an isolated soft ferromagnetic Fe nanoring with an inner diameter $d_{\mathrm{in}}=500$ nm, an outer diameter $d_{\mathrm{out}}=800$ nm and thickness $t = 80$ nm. Such dimensions allow for VS stabilization at remanence. To control the magnetization chirality we placed the NE, made also from Fe, inside the ring at distance of 25 nm from the inner wall of the NR. 
The shape of the NE can be regarded as a part of an NR with a width of 25 nm and sharp ends. The structure under investigation is shown in Fig.~\ref{fig:structure}(a). The shape of the NE gives shape anisotropy, with switching  field around 126 mT, which is higher than the coercive field of the NR at 100 mT.
In this paper, we compare 3 variants of the system: (1) non-NE configuration, i.e., the reference system, NR without the NE, (2) the parallel configuration with the external magnetic field parallel to the magnetization of the NE, and lastly (3) the anti-parallel configuration with the external field opposite to the magnetization of the NE.

All simulations were carried out using MuMax3, a GPU accelerated micromagnetic simulation program \cite{Vansteenkiste2014}.
To implement the system in the simulations, we discretized it with $512 \times 512 \times 7$ cells with cell size $\approx 1.57 \times 1.57 \times 11.42$ nm$^3$ for a total size of $805 \times 805 \times 80$ nm$^3$ along the \textit{x}, \textit{y} and \textit{z} axes, respectively. Because the magnetization is not uniform throughout the thickness and it would not be practical to show all 7 layers used in simulations, we will present in figure only the average magnetization across the thickness rather than a selected $z$-layer.

We used magnetic parameters from the experimental paper of Miyawaki, et al.~\cite{Miyawaki2006}. These are: the saturation magnetization \Ms = 1600 kA/m, the uni-axial magnetocrystalline anisotropy along the $z$-axis (out-of-plane direction) with constant $K = 47$ kJ/m$^3$, and the exchange stiffness constant \Aex~= 21 pJ/m.\cite{Miyawaki2006}
Using a Voronoi tessellation (see, Fig.~\ref{fig:structure}(b)), we have added magnetic grains in the NR to show that our results are robust even for imperfect materials. The grains are 20 nm on average and each of them has a random value of \Ms~obtained from a normal distribution of mean 1600 kA/m, and with a standard deviation \sigmams~of 2\% (32 kA/m). We also reduced the exchange coupling between grains uniformly by 5\% as compared to \Aex~value.
The introduction of this inhomogeneity results in a fully deterministic simulations for non-NE configuration, always leading to a single VS with CW or CCW chirality, for a given pseudo-random number defining grains and distribution of \Ms~among the grains.

The simulations for the statistical analysis (shown in Fig.~\ref{fig:stats}) for the parallel [Fig.~\ref{fig:stats}(b)] and non-NE configuration [Fig.~\ref{fig:stats}(a)] were run as follows. First, we apply a global external magnetic field \Bext~of 1 T along the $y$-axis, which saturates the NR and the NE if present, then we decrease \Bext~by 1 mT steps until we reach 0 mT. For every step, we use the conjugate gradient method to find the ground state of the magnetization. For the antiparallel configuration [Fig.~\ref{fig:stats}(c)], we could not apply such a strong external field as it would remagnetize the NE in the parallel configuration. So here, we start from \Bext~= 100 mT, which is sufficient to maintain the OS for the NR, but not enough to switch the NE. In this case, we also had to set the initial NE magnetization in the opposite direction to the external field to achieve the desired configuration. Then, the system was relaxed and finally we  demagnetized the rings to 0~mT in 1~mT steps. 

\section{\label{sec:results}Results and discussion}

\subsubsection{Chirality control demonstration}

\begin{figure}[ht]
\centering
\includegraphics[width=0.5\textwidth]{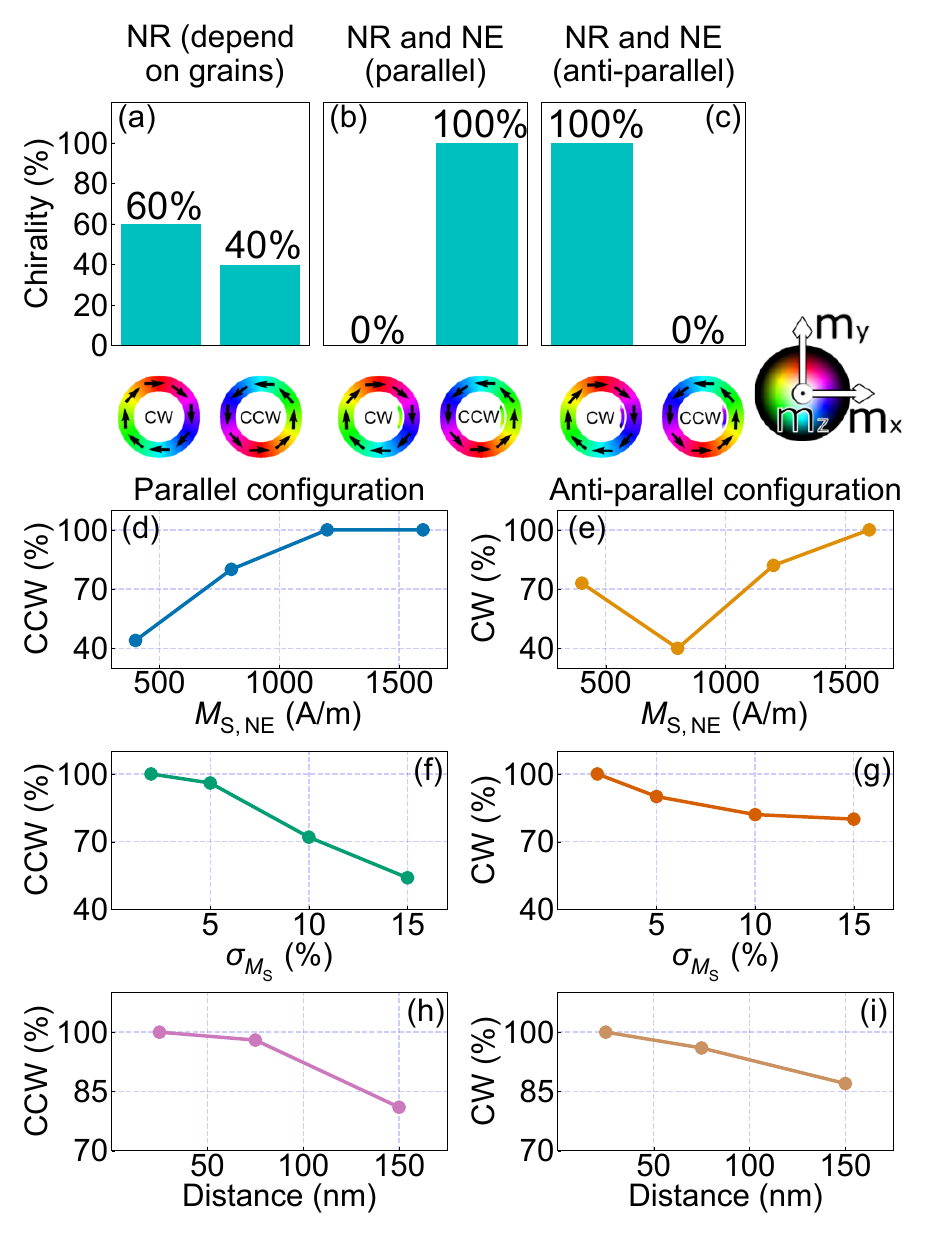}
\caption{\label{fig:stats}(a) The statistic of the 100 micromagnetic simulation results for the NR without the NE (non-NE configuration) with different random distribution of parameters among the grains. 
(b) and (c) The statistics of the micromagnetic simulations with the same grain realizations but for the NR with NE in parallel and anti-parallel configurations, respectively. At the bottom the static magnetization configurations in remanence are shown. Here, the color map indicates the vector orientation, where hue indicates the orientation of the magnetization according to the diagram in the right-bottom corner.
(d), (e) Statistic of the chirality control in dependence on the decrease of the magnetization saturation of NE. (f), (g) Chirality control depending on the increase in the distribution of magnetization saturation in grains. (h), (i) Chirality control in dependence on the separation between the NE and the inner edge of NR. }
\end{figure}

To show that the NE determines the final magnetization state of the NR, we conducted a statistical analysis of the remagnetization of the NR with decreasing the external magnetic field according to the procedure described in Sec.~\ref{sec:methods}. We ran 100 simulations for 3 configurations: non-NE configuration [Fig.~\ref{fig:stats}(a)], parallel [Fig.~\ref{fig:stats}(b)] and anti-parallel [Fig.~\ref{fig:stats}(c)]. For each simulation, we used a different random seed which resulted in a different organisation of the grain structure.

In Fig.~\ref{fig:stats}(a) we observe that the CW to CCW states are obtained in 60$\%$ and 40$\%$ of cases, respectively. The statistics change significantly when we introduce NE. From Fig.~\ref{fig:stats}(b) and (c) we see that we have full control of the VS chirality at remanence by the magnetization orientation of NE. 100\% of the simulations show a CCW configuration with a NE magnetized parallel to \Bext, and 100\% of the simulations show a CW configuration for the opposite case.

Figures~\ref{fig:stats}(d) and (e) show the effect of the magnetization saturation of the NE (\Msne)~on the chirality control for the parallel and anti-parallel configurations, respectively. We decreased the value of \Msne~from 1600 kA/m to 1200, 800 and 400 kA/m while keeping the value of \Ms~in the NR unchanged, i.e., at \Ms~= 1600 $\pm 5\%$ kA/m. We run 100 simulations in each case recording the final states. For both configurations we observe that lowering \Msne~reduces the degree of chirality control. This indicates that to control VS chirality the NE needs to have sufficiently strong magnetic moment, which can be guaranteed by sufficiently large \Ms~or volume of the NE.

Interestingly, for the anti-parallel configuration shown in Fig.~\ref{fig:stats}(e), the dependence is nonmonotonic. When the magnetization of NE is set to 800 kA/m, it causes the NE to switch its magnetization to be parallel to the nearest part of NR during the remagnetization process, resulting in the loss of chirality control. We observe that 40\% and 60\% of the VSs are CW and CCW, respectively (similarly to the non-NE configuration). However, when the magnetization of the NE is set to 400 kA/m, the in-plane magnetization changes to the out-of-plane magnetization. This is due to a perpendicular magnetic anisotropy which influence is enhanced with  reduced \Ms. This process   needs a separate investigations which are beyond the scope of this paper.

Figures~\ref{fig:stats}(f) and (g) show the statistics of the chirality in dependence on the \sigmams~distribution in the grains, which was determined using Voronoi tessellation. For both NE configurations, the effect of the degree of chirality control decrease as \sigmams increases. However, for antiparallel configuration this process is much slower in respect to the parallel configuration. Figs.~\ref{fig:stats}(h) and (i) show the statistics in dependence on the separation between NE and NR. With increasing distance, the effect of chirality control decreases for both configurations but up to 150 nm we still have over 80\% of control. This indicates that by selecting the position of the NE relative to the edge of the NR, we can tune the occurrence of chirality of a given type to a given probability.

The results demonstrate that we have a stable systematic control of the VS chirality using the NE during the remagnetization process. To elucidate the chirality control mechanisms, we performed a detailed hysteresis loop analysis.
 
\subsubsection{Remagnetization procedure}

NR and NE, made of the same material, exhibit different values of the switching field due to differences in shape anisotropy. This means that the magnetization reversal process of the NE will occur at different magnetic fields compared to the NR.\cite{Dong2021} Furthermore, it is important to note that the coupled NR-NE system significantly alters the value of the NE switching field,  it changes from \Hs~= 94 mT to \Hs~= 126 mT.

\begin{figure}[ht]
\centering
\includegraphics[width=0.5\textwidth]{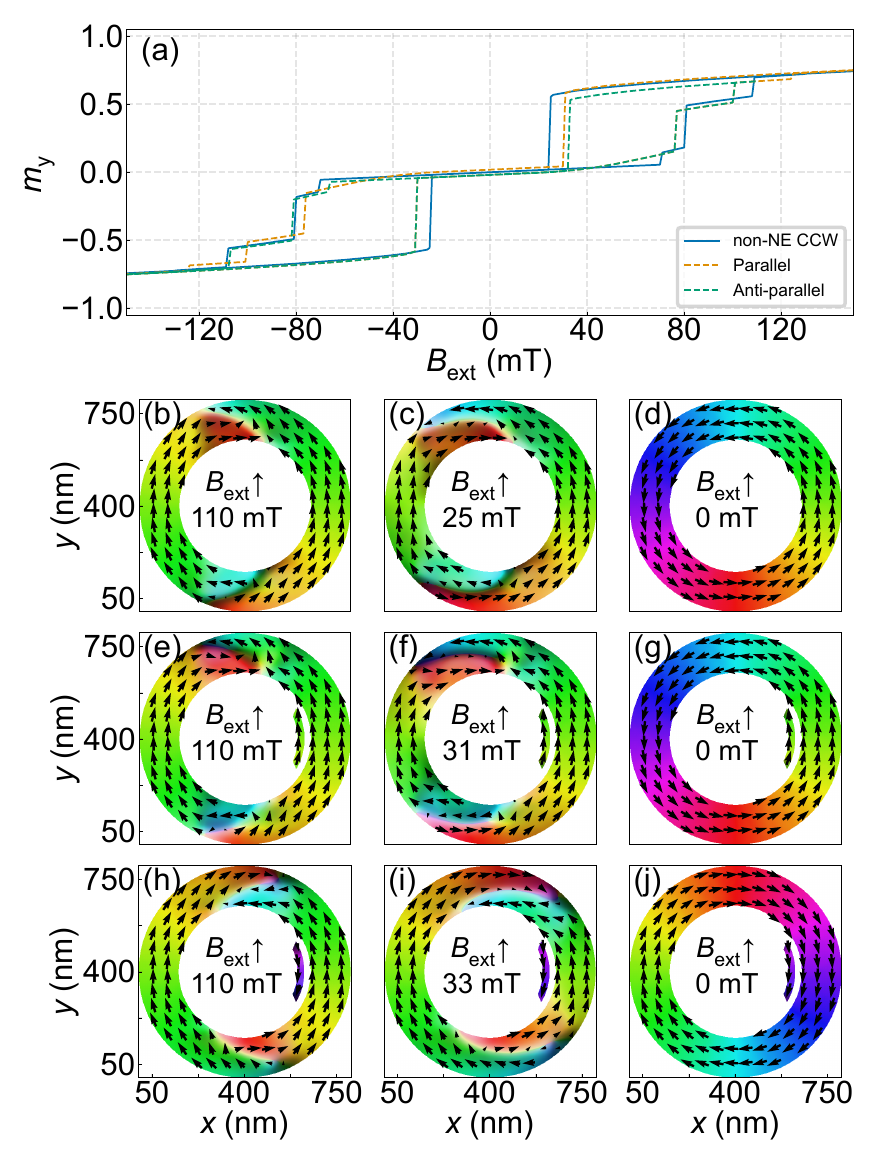}
\caption{\label{fig:hist} (a) The simulations of the hysteresis loop, where: for non-NE and parallel configuration, the simulations start at full saturation at -2000 mT to 2000 mT and opposite from 2000 mT to -2000 mT, for antiparallel configuration, the simulations start at full saturation at -2000 mT and interrupted at point 110 mT, then from 110 mT to -2000 mT. Magnetization configuration at selected magnetic fields: (b),(c),(d) -- in non-NE configuration, (e),(f),(g) -- parallel configuration, (h),(i),(j) -- antiparallel configuration. The colormap for magnetization orientation is the same as that shown on Fig.~\ref{fig:stats}.}
\end{figure}

We will analyse the hysteresis loop in the three scenarios related to the three configurations considered above.
In the first scenario, we consider NR in the absence of NE (non-NE configuration). The hysteresis loop is shown in Fig.~\ref{fig:hist}(a) with a blue solid line. We start simulations  by applying a large field of -2000 mT along the -$y$ direction, decreasing its magnitude to 0 mT, and then increasing it to the ring saturation in the opposite direction, i.e., 2000 mT. This process is then repeated by reversing the direction of the field. The HTH and TTT DWs appear when the magnitude of the field is less than 400 mT. As the field is decreased further, the DWs start to deform. The magnetization structures at 110 and 25 mT are shown in Fig.~\ref{fig:hist}(b) and (c), respectively. The remagnetization of the right part of the ring occurs at -24 mT, and a CCW state stabilises at remanence, as shown in Fig.~\ref{fig:hist}(d). However, the chirality of the ring depends on the random distribution of  the grains, as shown in Fig.~\ref{fig:stats}(a). Thus, for other grain distributions, a CW state also may stabilize.

The second scenario is for the NR-NE system in the parallel configuration, we also start with full saturation at -2000 mT. The hysteresis loop is shown in Fig.~\ref{fig:hist}(a) with a yellow dashed line, and it develops similarly to the previous scenario. Fig.~\ref{fig:hist}(e) shows the magnetization of the NR and NE at 110 mT. As shown in Fig.~\ref{fig:hist}(f), the magnetization of the system is just before switching from an OS to a VS at -30 mT. This switching occurs at a higher field than for the non-NE configuration. In this scenario, the NE is located on the right side of the NR, leading to a CCW magnetization chirality at remanence, as shown in Fig.~\ref{fig:hist}(g). Importantly, the final state in this scenario does not depend on the random distribution of  the grains, as demonstrated in Fig.~\ref{fig:stats}(b). Starting from a positive value of the external magnetic field, i.e., \Bext~= 2000 mT, we always reach a CCW chirality at remanence.

In the third scenario, we simulate a truncated hysteresis loop for the NR-NE system that allows for an antiparallel configuration, where the magnetization of the NE and the nearest side of the NR are aligned in opposite directions. We begin at an external magnetic field of -2000 mT along the $y$-direction and follow the main loop, decreasing the magnitude of the field as in the previous scenario. However, in this simulation, we interrupt the process at 110 mT [Fig.~\ref{fig:hist}(h)], when the magnetization of the NE and the nearest side of the NR are antiparallel. Then, we reverse the direction of the changes in the external magnetic field. Fig.~\ref{fig:hist}(i) shows the magnetization at 33 mT, just before the demagnetization of the NR. The magnetization switches in the part of the ring closest to the NE, establishing a CW chirality in the ring, as shown in Fig.~\ref{fig:hist}(j) and corresponding to Fig.~\ref{fig:stats}(c).

The results show that regardless of the direction of the external magnetic field at the starting point of the field change, parallel or antiparallel to the direction of magnetization in NE, the direction of magnetization in NE is the same as the direction of magnetization of the nearest side of NR at remanence, which determines the VS chirality.
Further conclusion is that the chirality control takes place at fields close to the switching field when one of the vertical parts of the ring changes the magnetization orientation as a result of DW motion in defined direction. This is clearly visible in the movies showing remagnetisation in NR provided in Supplementary Material. Thus, chirality is determined by the direction of movement of the DWs to the left or right part of the ring from the vertical symmetry axis, which is initiated by the magnetostatic interactions between NR and NE.

\subsubsection{Discussion}

\begin{figure}[ht]
\centering
\includegraphics[width=1.0\columnwidth]{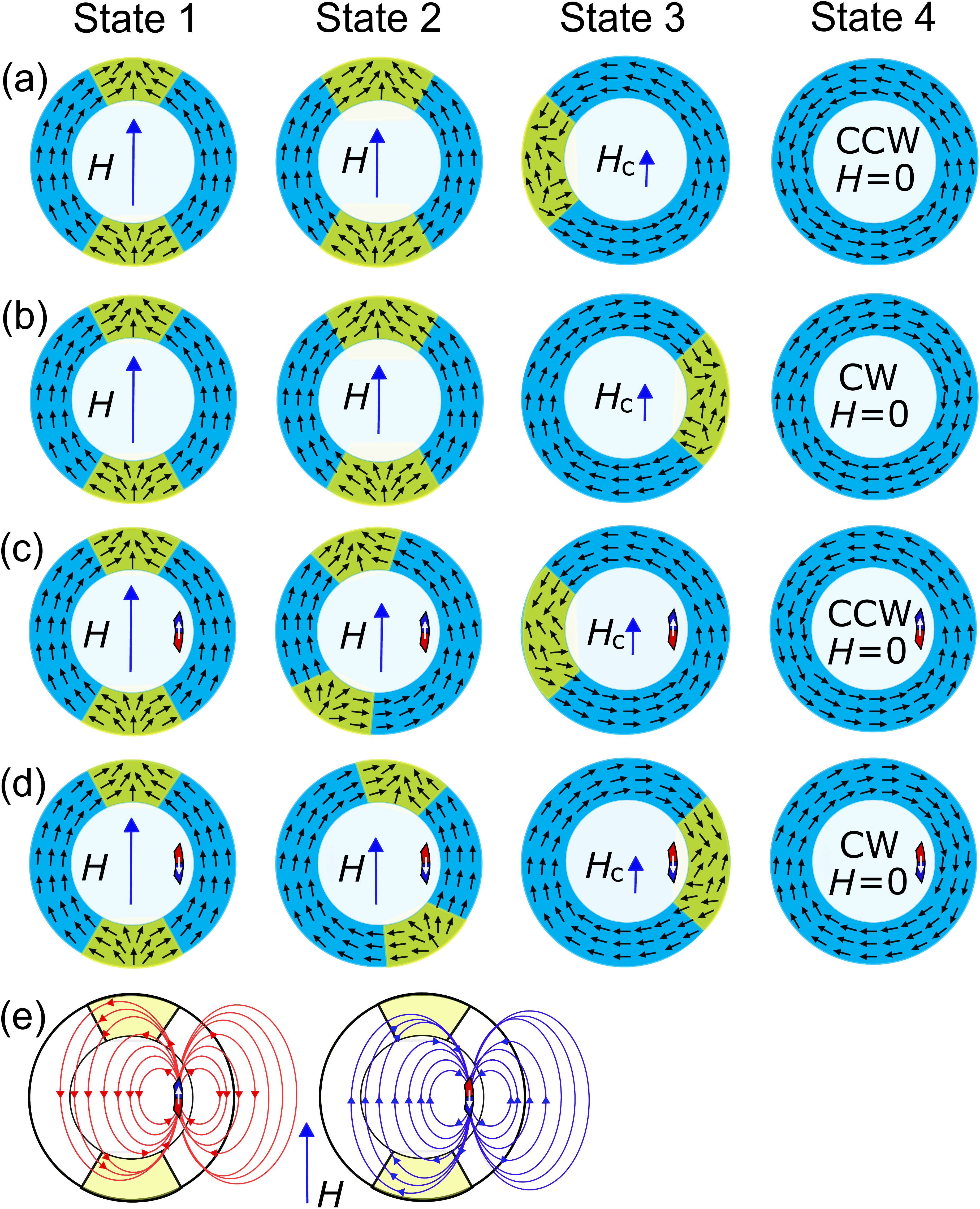}
\caption{\label{fig:magnetostatic} Schematic representation of the remagnetization process in a ring. Starting from the onion state (State 1), with decreasing magnetic field (applied vertically) the DWs move to the right or left (State 2) determining VS chirality at remanence (State 4) via annihilation at switching field (State 3). The chirality of the VS (State 4) is not controlled in the non-NE case (a), (b).  With the NE magnetization oriented parallel (c) or antiparllel (d) to the external magnetic field in a remanence the ring has determined the chirality, CCW and CW, respectively. (e) Schematic representation of the effect of the magnetostatic stray field from the NE on the DWs in NR.
}
\end{figure}

The stray field produced by NE interacts with the DWs and changes the internal field in the NR, so introduces the additional element to the system that control the direction of DWs propagation. 
In Fig.~\ref{fig:magnetostatic} we present schematically DW changes with decreasing magnetic field in the 3 scenarios of remagnetization discussed in the previous sections.

In the non-NE configuration, Figs.~\ref{fig:magnetostatic}(a) and (b), we can have two equivalent final configurations, CCW and CW, respectively. In the state 1 the DWs are in the HTH and TTT configuration. 
State 2 shows the DWs position at decreased field, where the DWs are placed with small tilts to the external magnetic field. The move of the DWs to the left or to the right are fully equivalent for perfectly symmetric NR, which leads to an uncontrolled VS chirality. States 3 and 4 show the process of annihilation of DWs and the final state, CCW or CW, respectively. 

Fig.~\ref{fig:magnetostatic}(c-d) schematically illustrates the stages of DW evolution during the remagnetization process for both parallel (c) and antiparallel (d) configurations. The process starts from State 1, where the ring has an onion state with HTH-TTT DWs in line with the external magnetic field. In State 2, the DWs begin to move, influenced by the magnetization of the NR. The NR induces a stray field, resulting in an asymmetrical distribution of the effective field between the left and right parts of the NR, as shown in Fig.~\ref{fig:magnetostatic}(e) and Fig.~S2(a-b) in Supplementary Material. For the parallel configuration of the NE and NR, the stray field from NE (left part of Fig.~\ref{fig:magnetostatic}(e)) pushes the DWs to the left side of the NR (State 3), ending with a CCW chirality in remanence (State 4). For the antiparallel configuration (right part of Fig.~\ref{fig:magnetostatic}(e)), the stray field from NE pushes DWs to the right arm of NR, causing the DWs to annihilate on the right side (State 3) and resulting in a CW chirality in remanence (State 4).

\begin{figure*}[t]
\centering
\includegraphics[width=\textwidth]{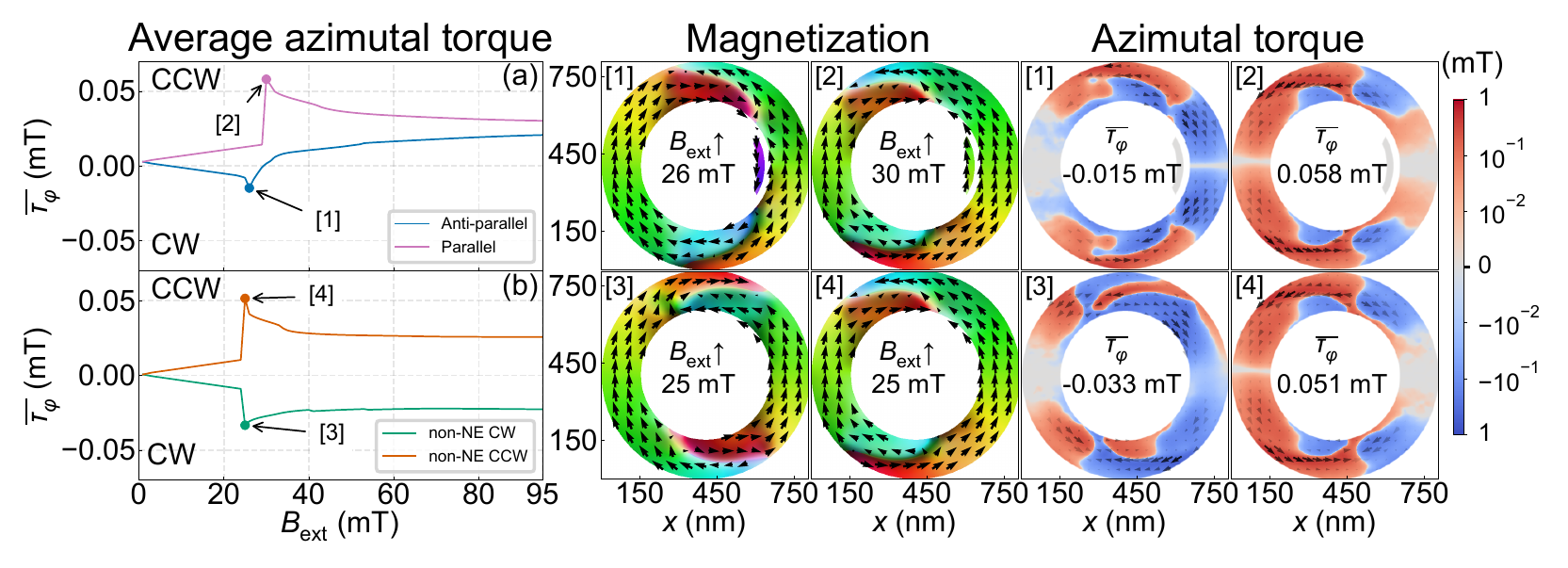}
\caption{\label{fig:torque} Numerical simulations of azimuthal component of the averaged torque in the ring in dependence on the magnetic field. In (a), we showed the results of two simulations for anti-parallel [curve (1)] and parallel [curve (2)] magnetization of the NE with respect to the magnetic field, which ends with CW and CCW VSs, respectively. In both simulations the same random seed was used. (b) The results for non-NE configuration with two different random seed numbers, which determine the chirality of the VS, CW (3) and CCW (4) in remanence. For (4) we used in simulations the same random seed as in (1) and (2). On the right part of the plots, the magnetization configuration and local distribution of the azimuthal component of the torque in NR are shown for four considered cases, at the fields just before the switching (marked by dots on the left plots).
The colormap for magnetization orientation is the same as that shown in Fig.~\ref{fig:stats}.
}
\end{figure*}

As described above, we brake the circular symmetry of the magnetic system by introducing NE, which creates a difference in the effective field distributions between the left and right parts of the NR and determines the direction of the DWs motion (see also Supplementary Materials) \cite{Klaui2003, Klaui2008}. Thus, the stray field produced by NE determines the direction of the torque exerted on the DWs in NR. 
To measure the effect of the NE on the NR, we extract from micromagnetic simulations magnetic torque for all the cases presented above.
We use the function, which is defined in Mumax3, and returns Cartesian components of the magnetic torque in (T) units. The torque is saved immediately after the change in the external magnetic field, before starting the relaxation procedure. We transform the torque for each discretization cell from the Cartesian to the 2D polar coordinate system. To obtain a single measure, we averaged the azimuthal component of the torque field across all spatial dimensions of the NR, resulting in a scalar, \ttorque, indicating whether the torque generated by the NE causes the NR's magnetization to rotate CW (\ttorque~$<$ 0) or CCW (\ttorque~$>$ 0).

In Fig.~\ref{fig:torque} (left parts) we present \ttorque~in dependence on the magnetic field for configurations considered above. For simulations of the curves (1), (2), and (4) we import magnetization texture for some selected seed from one of our previous analysis for non-NE configuration at the field 100 mT [Fig.~\ref{fig:stats}(a)], which ends with CCW VS. Then, we set manually the NE inside the ring with anti-parallel [curve (1)] and parallel [curve (2)] magnetization, and perform simulations with decreasing magnetic field, extracting \torque~at each field step. The simulation (3) in non-NE configuration, was performed for other random seed to show the remagnetization of the NR to the CW state. 
We see that the sign of \ttorque~at fields just before the switching field points out the VS chirality in remanence in all presented configurations. Moreover, for the case (1), where the magnetization orientation of NE was artificially reversed at the starting field 100 mT, the \ttorque~from the positive value changes the sign at smaller fields, and ends with negative values expected for CW chirality at this configuration.

\section{\label{sec:level1}Conclusions}

In summary, we have demonstrated with micromagnetic simulations the systematic control of the vortex chirality in a symmetric ferromagnetic ring by a  ferromagnetic nanoelement placed inside the ring. NE, by exerting stray magnetostatic field, changes symmetry of HTH-TTT DWs in the onion state, and during the remagnetization process determines the direction of the DWs movement and finally VS chirality in remanence. To control chirality NE requires sufficiently strong shape anisotropy to keep monodomain state and to have a switching field higher than the switching field of NR (without NE). In addition, the NE should have a sufficiently large magnetic moment (through large saturation magnetization or volume) to create a stray magnetostatic field that will determine the direction of DW motion. We show that this can be achieved by making NE with material in the shape similar to the part of the inner side of the ring, which simplify eventual fabrication process. In addition, we demonstrated the resistance of this method to the variability of geometric and material parameters of the system and random NR disturbances. All this make the experimental implementation of the proposed system using existing technologies possible and make it useful for spintronic and magnonic applications.

\begin{acknowledgments} 
The research that led to these results has received funding from the National Science Center of Poland, project no. 2020/37/B/ST3/03936.MM acknowledges funding from the Slovak Grant Agency APVV (grant number APVV-19-0311(RSWFA)), and supported by Research \& Innovation Operational Programme funded by the ERDF-ITMS project code 313021T081. The simulations were partially performed at the Poznan Supercomputing and Networking Center
(Grant No. 398).

\end{acknowledgments}

\nocite{*}

\bibliography{library}

\end{document}